# Carrying a Torch for Dust in Binary Star Systems

Daniel V. Cotton, Jonathan P. Marshall, Kimberly Bott, Lucyna Kedziora-Chudczer and Jeremy Bailey

*School of Physics, University of New South Wales, NSW 2052, Australia*

**Summary:** Young stars are frequently observed to host circumstellar disks, within which their attendant planetary systems are formed. Scattered light imaging of these proto-planetary disks reveals a rich variety of structures including spirals, gaps and clumps. Self-consistent modelling of both imaging and multi-wavelength photometry enables the best interpretation of the location and size distribution of disks' dust.

Epsilon Sagittarii is an unusual star system. It is a binary system with a B9.5III primary that is also believed to host a debris disk in an unstable configuration. Recent polarimetric measurements of the system with the High Precision Polarimetric Instrument (HIPPI) revealed an unexpectedly high fractional linear polarisation, one greater than the fractional infrared excess of the system. Here we develop a spectral energy distribution model for the system and use this as a basis for radiative transfer modelling of its polarisation with the RADMC-3D software package. The measured polarisation can be reproduced for grain sizes around 2.0 µm.

## Introduction

**Astronomical Polarimetry**

Within the Solar System polarimetry was most famously used to identify the Venusian clouds as being a composition of $H_2SO_4$-$H_2O$ particles of various sizes [1]. Venus is particularly well suited to this type of work because it presents a large range of phase angles to Earth in a short time frame, in contrast to the outer planets. Polarimetry is also used to learn more about particle size, composition and porosity of airless bodies like comets and asteroids [2]. Polarisation in these systems is phase and colour dependent with the maximum polarisation known to vary inversely with albedo [2]. Such determinations constitute complicated inverse value problems but are nevertheless pursued for the unique information they can provide.

The integrated light from the disc of a star is typically unpolarised; this is a result of the cancellation by circular symmetry of limb polarisation. For us to see polarisation from a star system requires that the circular symmetry be broken, or polarisation generated by scattering and absorption from surrounding gas and/or dust or another atmosphere. In this case it is possible to determine the size and composition of the scattering particles just as for Solar System bodies.

**The HIPPI survey of Southern bright stars**

We have recently completed and published [3] a short polarimetric survey of the brightest stars in the Southern hemisphere using the HIgh Precision Polarimetric Instrument (HIPPI) [4]. HIPPI is one of a new generation of parts-per-million polarimeters. The predecessor to HIPPI, PlanetPol [5], conducted a similar survey in the Northern hemisphere [6]. The unprecedented precision of these two instruments in fractional polarisation – more than an order of magnitude more precise than previous instruments used for such surveys – enabled the identification of a plethora of polarigenic phenomena.

Stellar types identified with elevated levels of polarisation included: late giants, Ap stars, Be stars, close binaries as well as ordinary B-type stars (particularly earlier types) [3], such that polarised stars were most prominent at the extremes of the H-R diagram. The signals from most main sequence stars are polarised only by the interstellar medium. The alignment of elongated grains within the interstellar medium by magnetic fields or otherwise acts to polarise light. Within the local hot bubble, a region largely devoid of gas and dust close to the Sun, this interstellar polarisation is small; ranging between $\sim 2\times 10^{-7}$ pc$^{-1}$ in parts of the Northern hemisphere to $\sim 3\times 10^{-6}$ pc$^{-1}$ at some Southern declinations [3] (although we have since tentatively identified a small localised region of the Southern sky with interstellar polarisation perhaps double this (Marshall et al., unpublished data), this doesn't appear to be the norm). Amongst main sequence stars, those known to harbour debris disks exhibited slightly elevated levels of polarisation.

Dust grains in circumstellar disks polarise light by scattering and absorption processes. Polarisation seen by aperture polarimetry – where the aperture takes in the central star as well as the whole/a large portion of the disk – has been reported at levels of $\sim 0.1$ to 2% ([7] and references therein).

**Debris Disks**

Circumstellar debris disks around evolved, main sequence stars are the dusty remnants of planet formation processes [8, 9]. They are composed of bodies spanning micron-sized dust grains to kilometre-sized asteroids. The larger bodies (beyond ~ mm-cm sizes) cannot be directly observed, but we infer their presence due to the short lifetime of dust grains under the influence of radiation forces [10, 11] in comparison to that of the host star. Often, the presence of circumstellar dust is revealed by the measurement of excess emission from a system, above that expected from the stellar photosphere alone, at infrared wavelengths.

From multi-wavelength modelling of the excess emission we derive a temperature (radial distance) for the dust [8], i.e. cool disks are detectable at long wavelengths, whereas warm/hot disks have signatures at increasingly shorter wavelengths. Broadly speaking, debris disks are either warm asteroid-belt analogues, or cool Edgeworth-Kuiper belt analogues (e.g. [12]). The incidence varies with spectral type; around FGK (sun-like) stars, the incidence of cool debris disks is 20+/-2% [13], and slightly higher for A stars at ~ 30% [14]. The observations are limited by instrumental sensitivity. At shorter wavelengths, where the photosphere is a greater contribution to the total emission, the contrast between star and disk drops such that only 2% of FGK stars are known to host disks [15]. Likewise, determining the presence of a faint warm belt in the presence of a bright cool belt is also tricky [16].

 The architectures of debris systems derived solely from the excess are subject to inherent degeneracies in the modelling process (e.g. between dust grain size and temperature). For example, a simple blackbody approximation underestimates the disk extent for FGK stars by a factor of 4-5 (e.g. [17, 18]).  Using spatially resolved imaging as an additional constraint weakens the degeneracy and allows a much better constraint on the dust properties to be derived [19, 20]. Images of the disk continuum at infrared wavelengths are limited by the angular resolution offered by space telescopes. Imaging a disk in scattered light with ground-based 10m-class telescopes (or interferometers) provides both the high spatial resolution imaging, and a measure of the scattering properties of the dust grains - information that cannot be gleaned by alternative means. However, most debris disks are tenuous, and the relationship between continuum excess and scattered light brightness is weak [21].

Polarimetric imaging using the high-contrast adaptive optics instrument GPI has provided new insights into the structure and architecture of several systems [22-25]. Tracing the polarisation of the disk as a function of its radial extent and orientation allows a unique determination of the disk alignment, and greater insight into the dust properties throughout the disk rather than in aggregate. Aperture polarimetry with e.g. HIPPI [4] can provide similar or even better sensitivity to polarisation induced by circumstellar dust, but without the spatial information, useful in cases where this would be unobtainable in any case e.g. for asteroid belts around more distant stars, or 'hot dust' disks.

**Epsilon Sagittarii**

*System*

The inspiration for the present paper is ε Sgr (HIP 90185); it is unusual for a debris disk system on two counts, firstly it has a spectral type of B9.5III, and secondly it is a binary system where the secondary is separated from the primary on a similar scale to the debris disk. It is this second property that makes ε Sgr particularly interesting, for it has the potential to produce an asymmetry in the scattered light seen in our aperture and thus a polarisation signal beyond what might be expected from the disk otherwise.

ε Sgr A is 3.52 $M_{Sun}$ star of spectral type B9.5III. The secondary, ε Sgr B, has a mass of 0.95 $M_{Sun}$ [26] and orbits at 106 AU from the primary [27]. Rhee et al. [28] have reported an excess of $4.5 \times 10^{-6}$ from *IRAS* 60 μm data, indicating a debris disk centred at 155 AU [27]. Consequently Rodriguez and Zuckerman [27] list ε Sgr as one of nine known dynamically unstable binary/multiple debris disk systems, with only a 0.2% probability that uncertainties in inclination and orbit actually place the secondary outside the unstable zone. The excesses derived from *Spitzer* data at 13 μm and 31 μm [29, 30] are greater than that at 60 μm from *IRAS*, which implies a closer debris disk. Below we develop a spectral energy distribution model with this and other data. Both the cited references and our model assume a circumprimary debris disk, however it should be mentioned that a circumsecondary system like that described by Rodigas et al. [31] for HD 142527 may also be a possibility. Regardless of the true nature of the system we feel that the assumed geometry for ε Sgr represents an interesting case for investigation of the effects of a secondary on the observable features of a debris disk system.

*Polarimetry*

A linear polarisation measurement of $162.9 \pm 4.4 \times 10^{-6}$ for ε Sgr was obtained with the HIPPI instrument operating in the SDSS g' band (effective wavelength 462 nm) on 1/9/2014 [3]. HIPPI has an aperture of 6.7", which would place the secondary within the aperture, and the centre of the debris disk on the edge of it. However, seeing at the time of the observation was ~4", meaning that a significant contribution from the outer part of the disk can be counted on.

Based on the maximum interstellar polarisation we observed during the bright star survey we would not expect more interstellar polarisation than $\sim 132 \times 10^{-6}$ for a star at 44 pc in the South, meaning that at least $\sim 30 \times 10^{-6}$ is intrinsic to the system – more than would be expected from the infrared excess (Marshall et al., in prep). In the particular case of ε Sgr, the nearest star that we've measured that we expect to have only interstellar polarisation is the A2.5V star ζ Sgr. It has an interstellar polarisation corresponding to $1.03 \times 10^{-6}$ pc$^{-1}$ [3]. Assuming a linear increase in polarisation with distance this suggests an interstellar contribution of only $45 \times 10^{-6}$ for ε Sgr. However ζ Sgr and ε Sgr are separated by 10 degrees and 17.9 pc, and even though the magnitude of interstellar polarisation appears to vary fairly smoothly near to the Sun this calculation should be considered as potentially indicative only.

It should also be noted that ε Sgr B was found as a result of a search of late-B stars showing high X-ray fluxes [26]. Some X-ray binaries have been found to show variable polarisation [32] which may be an alternative explanation for the polarisation observed. However, such detections have been rare, and generally for much stronger X-ray sources. Indeed the X-ray activity might be an indication of dust accretion onto the secondary, which is another scenario suggested for HD 142527 [31].

## Modelling Procedure and Results

**Spectral Energy Distribution**

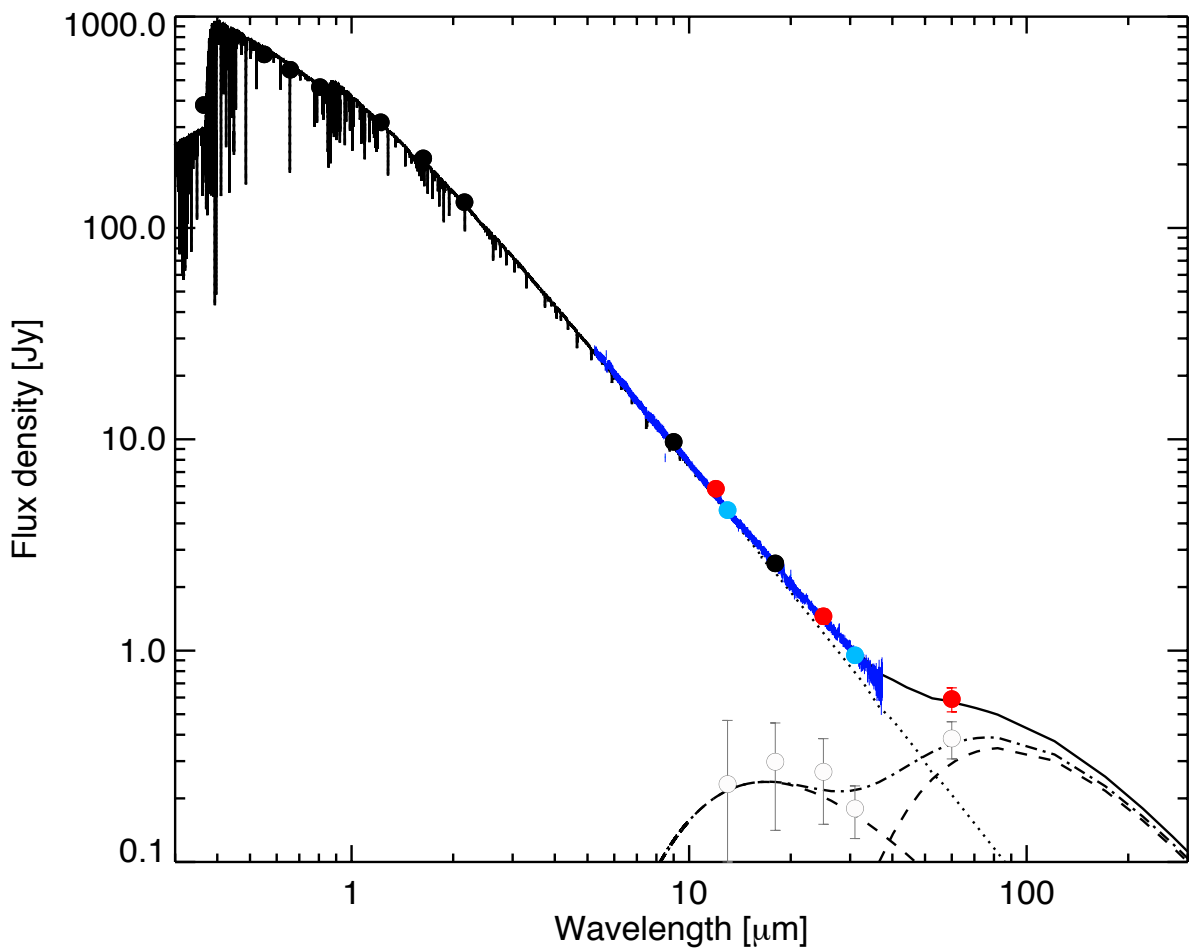

*Fig 1: ε Sgr SED. Black data points are ancillary optical, near- and mid-infrared photometry. The dark blue line is the Spitzer IRS spectrum, and the light blue data points are fluxes derived from the spectrum at 13 and 31 μm. Red data points are IRAS photometry. Hollow data points denote the excess at mid- and far-infrared wavelengths. 1-sigma uncertainties are shown as error bars. The dotted line is the stellar photosphere, combining both primary and secondary components, the dashed lines are the disk components, the dot-dash line is the sum of the disk components, and the solid line is the total star + disk model.*

Fig 1 shows the modelled spectral energy distribution (SED). We modelled the spectral energy distribution of the system using two blackbody components and obtained a fractional excess of $75 \times 10^{-6}$, which is more consistent with what we would expect based on the polarisation measured. The quality of this value is strongly dependent on the assumptions made regarding the temperature of the cold component and the stellar photosphere contribution.

The contribution of the stellar binary to the total emission was represented by a pair of NEXTGEN stellar atmosphere models [33, 34]. The primary (B9.5III) was modelled with $T_{eff}$ = 9800 K, log g = 3.5, and R* = 3.7 Rsol, whilst the secondary (K5) was modelled with $T_{eff}$ = 5600 K, log g = 4.5, and R* = 1.1 Rsol. The stellar radii are derived from a least squares fit to the photometry based on the stellar models used, which are appropriate for the spectral types. Both stars were assumed to have solar metallicity. The combined stellar SEDs were scaled to optical and near-infrared photometry taken from *HIPPARCOS* [35], 2MASS [36], and WISE [37].

The disk model was approximated as a pair of blackbodies. This approach was motivated by the rising slope of the IRS spectrum at mid-infrared wavelengths combined with the strong excess at 60 μm, indicating the possible presence of two dust components. In our model the two dust components are fitted simultaneously using a least-squares approach, weighted by the photometric uncertainties. We find the best-fit temperature for the cold component, fitted to the 31 and 60 μm data points, to be 60±10 K ($L_{ir}/L_* = 7.4 \times 10^{-5}$), and the warm component, fitted to the mid-infrared excess, was found to have a best-fit temperature of 300±50 K ($L_{ir}/L_* = 1.3 \times 10^{-6}$). We note that a single, warmer disk model does not replicate the shape of the IRS spectrum and IRAS 60 μm data satisfactorily. Due to the absence of longer wavelength data constraining the peak of the cold emission, the total fractional excess is subject to large uncertainties.

**Radiative Transfer and Polarisation**

To estimate the disk polarisation we use RADMC-3D [38], a three-dimensional radiative transfer modelling code, to handle the scattering and polarisation of a given dust composition and disk structure. We assume that the disk architecture is a ring of uniform volume density of dust centred 120 AU from ε Sgr A with a diameter of 20 AU. The ring is wedge shaped with a narrow opening angle of 5 deg. The geometry of the system is shown in Fig 2. The dust composition used in our models is astronomical silicate [39]. The SED model used blackbodies to fit the excess but here we adopt a realistic grain model. To reconcile these two approaches, the emission from the RADMC-3D model, using the architecture inferred from the SED model, is scaled by an arbitrary amount to match the observed excess emission i.e. the dust mass, number density, etc. is scaled to achieve the same excess.

Using the DDSCAT code [40] we calculate the Mueller matrix elements for an ensemble model of the dust assuming a single grain size within the disk. A power law distribution would be consistent with a steady-state collisional cascade [41], and such is an approach we intend to take with further work, but here we are predominantly interested in the potential scale of the effect and so have taken a simpler approach for the sake of expediency. These matrix elements, along with the absorption and scattering efficiencies are then used as input to RADMC-3D to compute the disk model. Output from the disk model consists of four images tracing the Stokes parameters *I*, *Q*, *U* and *V* at 550 nm (which is the centre of the standard Johnson V band), of which we present *Q* and *U*. The models below in Fig 3 through Fig 5 illustrate the expected polarisation from ε Sgr as a function of system inclination, geometry and dust grain size. Note that the RADMC-3D polarisation co-ordinate system is rotated 90 degrees such that positive Q aligns with the x-axis.

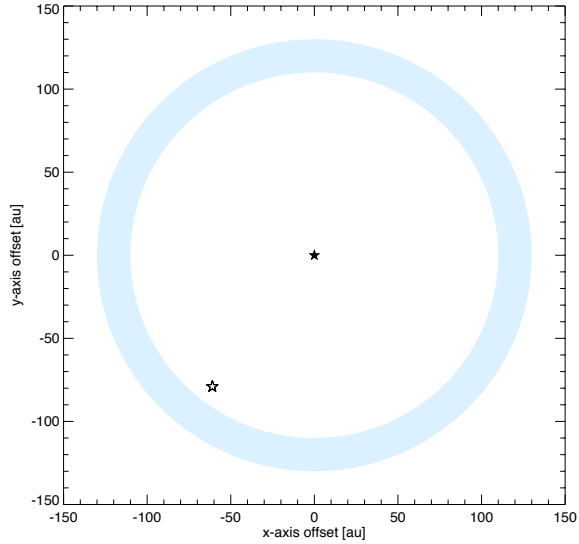

*Fig 2: A schematic layout of the ε Sgr system as modelled. We have assumed a narrow, 20 AU, annulus for the disk in addition to known system parameters. In the latter polarisation maps the x- and y-axis units of 1 to 100 correspond to the -150 to 150 AU range shown here.*

*Table 1: Integrated polarisation for each geometry and grain size modelled. (0.1 μm sized grains were also tested but with no appreciable polarisation signal in any geometry.)*

| Particle Size (μm) | ε Sgr B PA (deg) | Inclination (deg) | Q (x$10^{-6}$) | U (x$10^{-6}$) | P (x$10^{-6}$) |
|---|---|---|---|---|---|
| 0.5 (Small) | 0 | 0 | -0.4 | 0.0 | **0.4** |
|  |  | 60 | 0.6 | 0.0 | **0.6** |
|  |  | 90 | 1.0 | 0.0 | **1.0** |
|  | 90 | 0 | 0.4 | 0.0 | **0.4** |
|  |  | 60 | 1.6 | 1.1 | **1.9** |
|  |  | 90 | 2.5 | 0.1 | **2.5** |
| 1.0 (Medium) | 0 | 0 | -0.2 | -0.1 | **0.2** |
|  |  | 60 | -12.0 | 0.0 | **12.0** |
|  |  | 90 | -12.3 | 0.0 | **12.3** |
|  | 90 | 0 | 0.0 | 0.1 | **0.1** |
|  |  | 60 | -14.3 | -4.2 | **14.9** |
|  |  | 90 | -15.2 | 0.6 | **15.2** |
| 2.0 (Large) | 0 | 0 | -16.0 | -0.6 | **16.0** |
|  |  | 60 | -12.0 | -0.1 | **12.0** |
|  |  | 90 | -12.3 | 0.0 | **12.3** |
|  | 90 | 0 | 16.0 | -0.2 | **16.0** |
|  |  | 60 | -110.0 | -39.0 | **116.7** |
|  |  | 90 | -200.0 | 71.0 | **212.2** |

For inclinations other than 0 degrees, the position angle of the secondary is affecting the polarisation that we see, this can be seen by comparison of Fig 3 and Fig 4. Here greater polarisations are generated for ε Sgr B at a position angle (PA) of 90 degrees. This is a manifestation of asymmetry of the system from the perspective of the observer. Here we also see more significant polarisations in the *U* Stokes parameter. Again this is a result of asymmetry, but here it is brought on by the directionality of Mie scattering, i.e. there is greater forward scattering for the larger grains, and when the disk is inclined this manifests in *U*.

The single grain sizes tested here are on the same scale as the wavelength of light. Consequently there is a "ringing" effect superimposed on trends with grain size. Nevertheless, in Fig 5 we can see the result of increased forward scattering with increased grain size where the polarisation in *Q* near the secondary goes from mostly positive to mostly negative.

The most striking feature of Fig 3 through Fig 5 is the impact of the secondary on the polarisation map – intense fractional polarisation is localised near the secondary. Table 1 presents the total polarisation for the system in each geometry and for each grain size tested. It shows that the total polarisation is consistently larger when the secondary is off-centre and an asymmetry induced in the integrated system. Furthermore, Table 1 reveals that for the largest grains tested the polarisation produced is comparable to the polarisation measured by HIPPI for ε Sgr, and is greater than the infrared excess of the system.

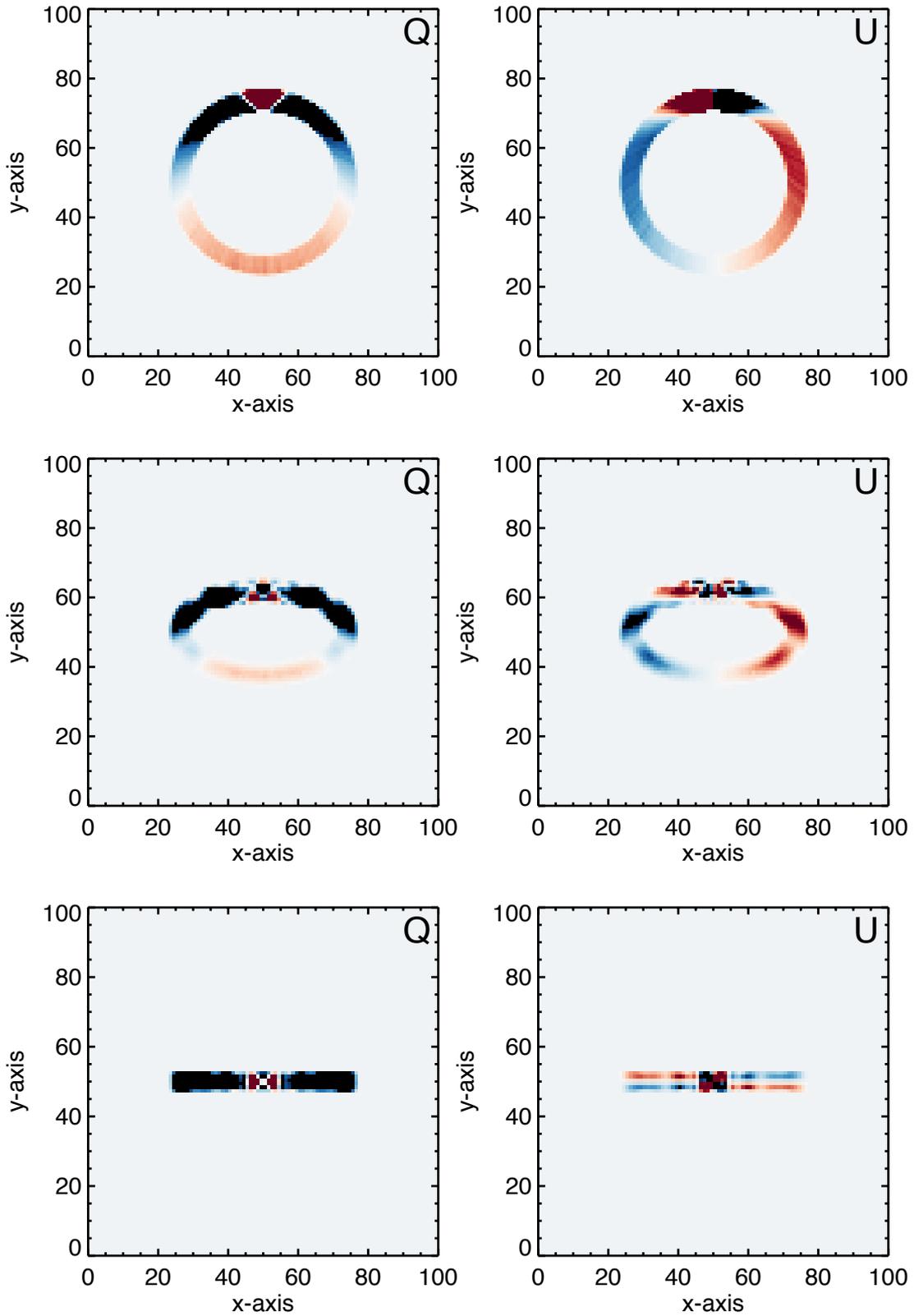

*Fig 3: Polarisation maps (fractional Q and U) of the ε Sgr disk at three inclinations (0, 60 and 90 deg, top to bottom) and for 1 μm sized dust grains. The secondary is positioned at 0 deg in the viewing plane. The magnitude of the colour scale is arbitrary but runs from positive red through white to negative blue, where negative Q is N-S on the diagram.*

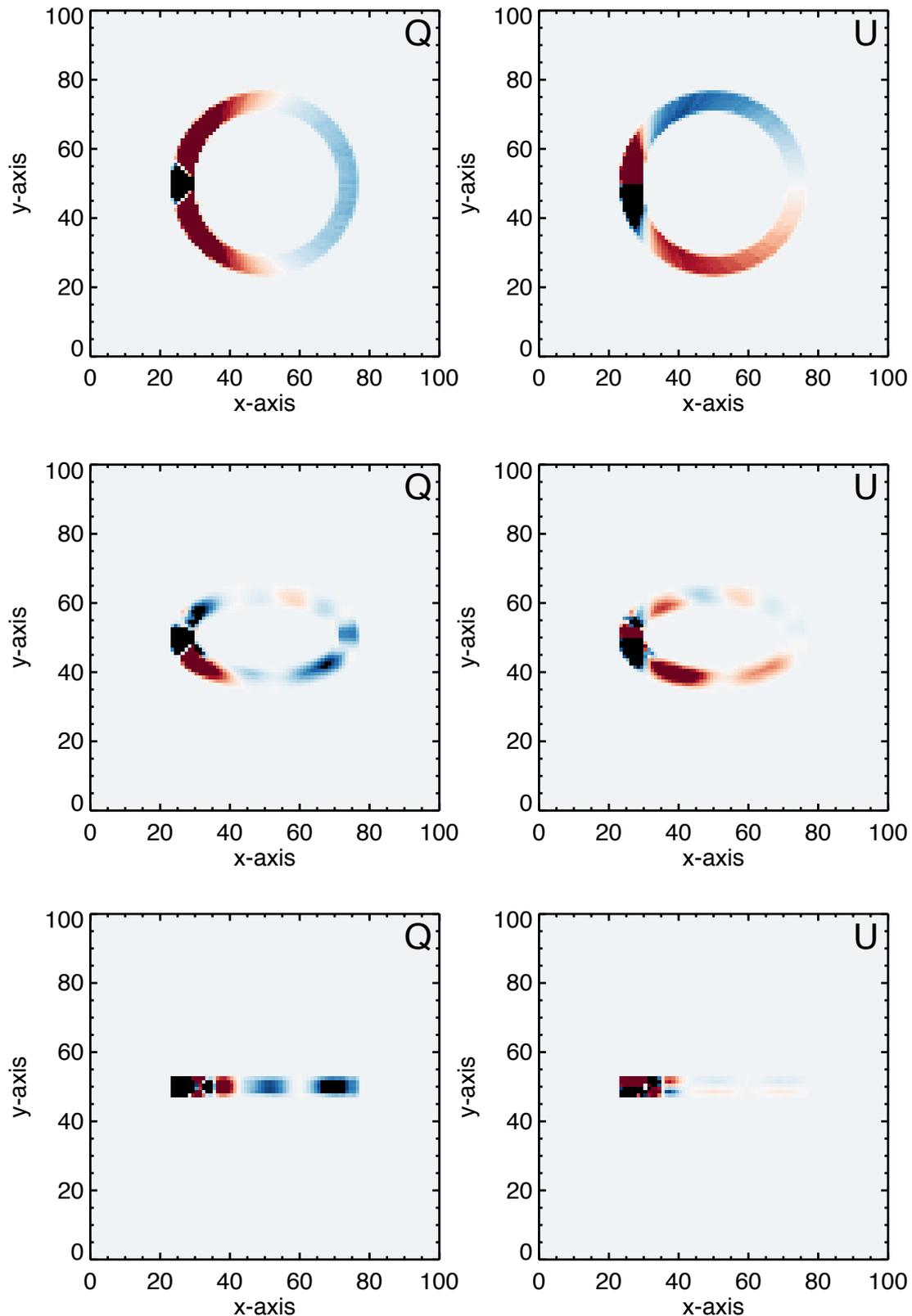

*Fig 4: Polarisation maps (fractional Q and U) of the ε Sgr disk at three inclinations (0, 60 and 90 deg, top to bottom) and for 1 μm sized dust grains. The secondary is positioned at 90 deg in the viewing plane. The magnitude of the colour scale is arbitrary but runs from positive red through white to negative blue, where negative Q is N-S on the diagram.*

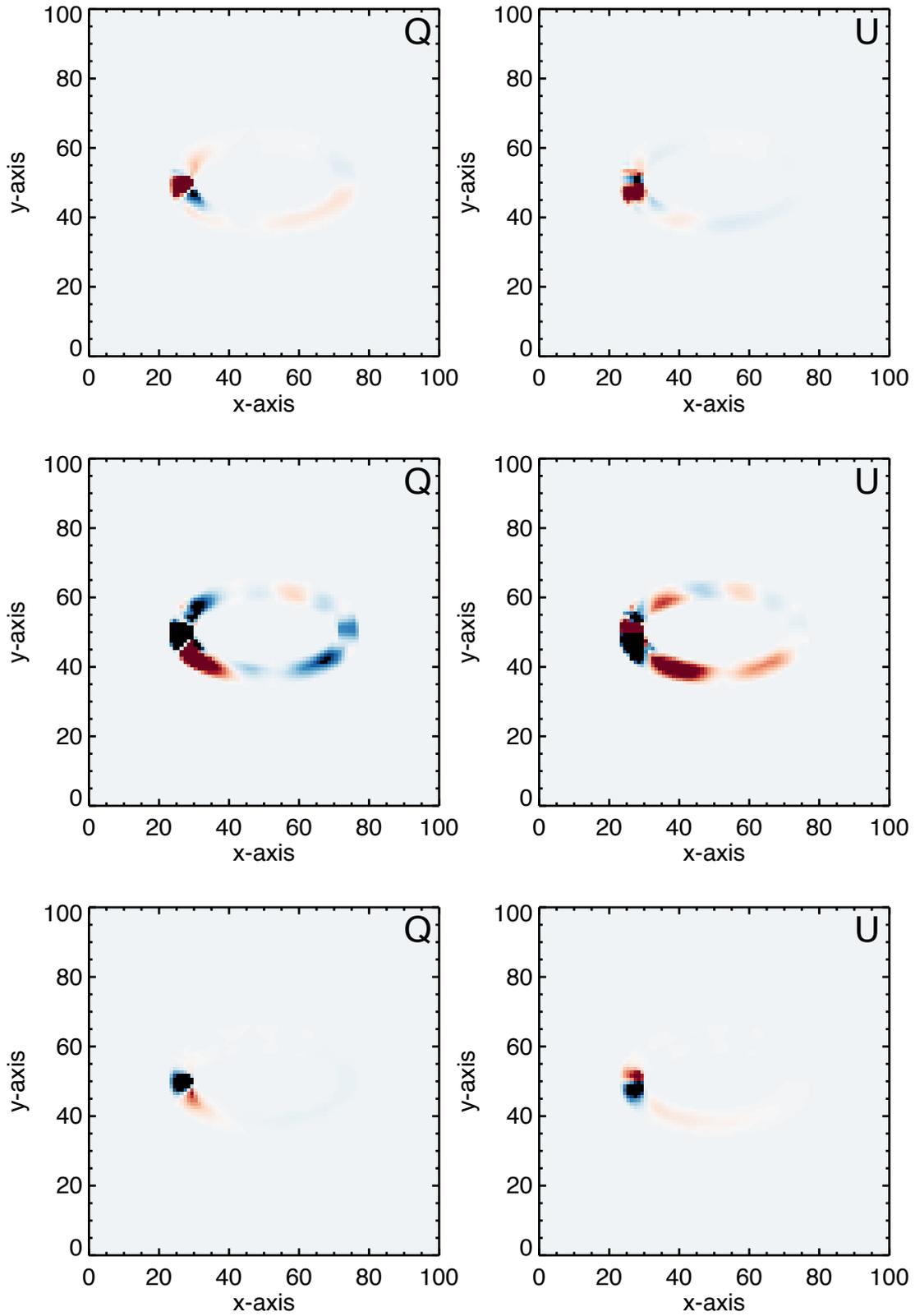

*Fig 5: Polarisation maps (fractional Q and U) of the ε Sgr disk at 60 deg inclination for three different dust grain sizes (0.5, 1.0 and 2.0 μm, top to bottom). The secondary is positioned at 90 deg in the viewing plane. The colour scale presented here is qualitative, with the scale for the large grains reduced by a factor of 100 compared to the small and medium sized grains to accommodate a larger dynamic range.*

## Discussion and Conclusions

The calculations performed here are simplistic and may not accurately reflect the detailed geometry of the ε Sgr system. We have modelled only spherical grains, whereas real dust grains are irregular in size and thus would generate larger degrees of polarisation. Additionally, the (detectable) grains in the disk will not be single-sized but will have a distribution of sizes from the blow-out radius up to mm. The *blow-out radius* – the radius of the largest grains blown out of the system by a stellar wind – is a function of the stellar luminosity, with a larger blow-out radius for dust around more luminous stars [11]. For ε Sgr B the blow out radius might be as small as 0.5 μm, but as ε Sgr A is a giant and close we'd expect the smallest grains near the secondary to be larger than that. For a system such as this, with a 4 $L_{Sun}$ primary, ~ 10 μm would be a typical size. The grain sizes tested are therefore fairly realistic. Smaller grains may be present in the system, produced in on-going collisions, but will be swiftly removed. It is noted by Rodriguez and Zuckerman [27] that the disk is unlikely to be stable given the architecture they infer from the available photometry, and perturbation of the disk by the secondary star would be one source of dynamical excitement driving collisions between bodies within the disk. A full model of the dynamical stability of circumbinary or circumsecondary disk architectures is beyond the scope of this work.

The larger grain sizes we modelled produced a level of polarisation greater than the infrared excess of the system. The magnitude of the polarisation produced being similar to what was measured for the system. The position angle of the system was measured in March 1999 to be 142.3 [26]. The measured polarisation angle was 38.1±0.8, which, neglecting the uncertainty, is a difference of 104.2 degrees. The polarisation produced by the models in the more favourable geometry for the larger grains is mostly negative *Q*, which is 90 degrees different to the secondary PA. Our modelling suggests the deviation from 90 degrees could be due to the inclination of the system, but it could also be a result of motion of the secondary since 1999 – if the system is face on (0 degree inclination) this could be as much as 11 degrees. This shows us that the large degree of polarisation measured could plausibly be due simply to the debris disk's interaction with the binary star system, without the need to invoke other polarisation mechanisms.

## Further Work

The disk here is assumed to be smooth and radially symmetric. Future work will include an investigation of the effect of the addition of one or more clumps into the disk. Should such structure produce a significant effect then there is potential for using the secondary as a torch to illuminate progressive sections of the debris disk. Such a technique could be useful for more distant systems where imaging polarimetry is not possible.

ε Sgr B is a redder star than ε Sgr A, we would therefore expect a stronger polarisation signal at redder wavelengths. However, effects to do with the grain size distribution may offset this. In addition to looking at more realistic size distributions we also intend to carry out the calculations in a greater range of wavelengths and probe the ε Sgr system with multi-band polarimetry to inform and test the modelling.